\documentclass[11pt,twoside]{article}

\usepackage{asp2006}
\usepackage{epsf}
\usepackage{lscape}

\begin{document}
\title{On the Road to Understanding Type Ia Progenitors: Precision
Simulations of Double Degenerate Mergers} 
\author{Chris L. Fryer, Steven Diehl} 
\affil{Los Alamos National Laboratory, Los Alamos, NM 87545} 
\affil{C.L.F. also at University of Arizona, Tucson, AZ 85721}

\begin{abstract} 
We review the current state of the art in double degenerate merger
simulations to better understand the role this phenomenon plays in
type Ia progenitors.  Because the fate of a merged system may well
depend on the exact evolution of the matter temperature (as well as
mixing of the merged system), precision simulations are required to
determine the true fate of these systems.  Unfortunately, if we
compare the results of current simulations, we find many-order of
magnitude differences in quantities like mass-transfer rates in the
merger process.  We discuss these differences and outline an approach
using verification and validation that should allow us to achieve a
level of precision sufficient to determine the true fate
(thermonuclear vs. collapse) of double degenerate mergers.
Understanding the fate of lower-mass systems (e.g. R Coronae Borealis
stars) may be key in our final testing phase.

\end{abstract}


\section{Introduction}

Supernovae (SNe) and their Gamma-Ray Bursts cousins mark the most
powerful explosions in the universe.  The quest for an understanding
of the driving mechanism behind these explosions has been nearly as
turbulent and heated as the events themselves.  It is now nearly
universally agreed that these explosions are produced in one of two
engines: the collapse of a massive stellar core (the gravitational
potential energy released in the collapse being the energy source of
the supernova explosion), the thermonuclear explosion of a white dwarf
(where the fusion of carbon and oxygen into heavier elements releases
the energy to power the explosion).  Nearly as fractious has been the
discussion of the progenitors of these mechanisms.  Understanding the
progenitors of supernovae requires an understanding of stellar
evolution, something we are far from doing.  For ``core-collapse''
supernovae, SN 1987A provided our first direct glimpse of a supernova
progenitor because the progenitor was observed prior to the explosion
(Kirshner et al. 1987).  It also profices an example of how wrong
stellar evolution can be (past models insisted that only red
supergiant stars would collapse to form supernovae).  But for
thermonuclear explosions (type Ia Supernovae), we have not been so
lucky.  All current type Ia SNe progenitors require mass transfer from
a binary companion.  To understand these progenitors, we must not only
understand stellar evolution, but also binary interactions,
introducing an entirely new set of uncertainties into our
understanding of the type Ia progenitor.

One of the persistent progenitor scenarios of type Ia supernovae is
the merger of two white dwarfs (double degenerate scenario), producing
a single white dwarf with mass above the Chandrasekhar limit that will
contract and explode.  The primary drawback of this scenario is that
current theory argues that such a merger will not produce a type Ia
supernovae.  But its advantages, mostly that theory predicts that the
rate of such mergers is consistent with the supernova rate (for many
other proposed scenarios, this is not the case), has kept this
proposed progenitor alive.  To understand why this scenario is not
believed to work, we must understand what happens when a white dwarf
accretes matter.  Nomoto and Kondo (1991) summarized this
understanding in a single plot (Fig. 4 of that paper).  They
determined the fate of a Carbon/Oxygen (CO) white dwarf as a function
of its birth mass and the rate at which we accrete material on this
white dwarf.  For the double degenerate scenario, the region of
concern is at the topmost accretion rates.  When the accretion rate is
above a few times 10$^{-6} {\rm M_\odot s^{-1}}$, the carbon in the
white dwarf ignites at its edge and burns inward, transforming the CO
white dwarf into an Oxygen/Magnesium/Neon white dwarf.  When such a
white dwarf approaches the Chandrasekhar limit and contracts, neutrino
emission cools the white dwarf sufficiently to prevent a thermonuclear
runaway until the matter has contracted too much for the explosion to
escape its own potential well.  This material will continue to
collapse until nuclear forces and neutron degeneracy halt the collapse
at the formation of a neutron star.  The fate of such an object is
interesting (it is termed Accretion Induced Collapse ``AIC'' and has
been used to explain a number of neutron star populations), but
definitely not a type Ia supernova (Fryer et al. 1999).  As we shall
discuss below, all current results showing the merger of 2 white
dwarfs suggest that this process is rapid and that the ultimate
accretion rate onto the white dwarf will be very close to the
Eddington rate.  Hence, theory currently predicts that the double
degenerate scenario produces AICs and not type Ia supernovae.

There are a number of caveats to this result.  The Nomoto \& Kondo
result assumed constant accretion rates and did not account for the
fact that the white dwarf could have a very complex rotation profile
(Saio \& Nomoto 2004; Yoon et al. 2007).  This has led to small
windows of opportunity for the double degenerate scenario to still
produce type Ia supernovae.  But to determine what the true fate of 
these objects is, we must simulate the merger.  And we must be able 
to believe the results of our merger in detail.

This brings us to an important concept in the scientific method.  In
astronomy, we have two types of theoretical investigations: predictive
and, for symmetry sake, `` post-dictive''.  Predictive science is what
we strive for - to be able from first principles to determine how
something should behave.  Post-dictive is what happens with a lot of
science.  We know the answer we must get (e.g. explaining the solar
abundance pattern) and we fit in some free parameters in our model to
make sure to get this answer.  Of course, we can not predict errors in
the observations, and when the solar abundance pattern changes, we
must then simply accept this change.  Post-dictive science has many
virtues: first, if we can match data with a reasonable set of
parameters, we show that our basic model may also be reasonable;
second, the parameters required for such a model to fit the data might
teach us something about the underlying physics requirements.
Unfortunately, scientists often forget that they have made parameter
assumptions and start to believe they have predicted the answer.  In
such scenarios, post-dictive models can do more harm than good
(scientists may argue they have solved a problem prematurely,
preventing continued work on a subject).  Stellar evolution is rife
with examples of such misuse of post-dictive models.

But truly predictive models are very difficult to do in astrophysics.
Generally the problems we are interested in astrophysics are too
complex to be solved with a simple analytic derivation.  Once we
resort to numerical models, we must be wary about numerical artifacts
in our simulations.  The national laboratories have focused their
testing procedure on a process of Verification and Validation (V\&V).
Verification is the process by which scientists test to make sure their
numerical models are solving correctly the physics in their code.
Validation is the process by which scientists confirm that the physics
in their code is the correct physics for the problem they are solving.
Most of our time is spent on Verification, which can take many forms:
comparison to analytic solutions, convergence studies, code
comparison, and even comparison to laboratory experiments.  Validation
is almost entirely focused on comparison to some observation or
experiment.

In this paper, we review the current status of simulations of WD
mergers.  Recent simulations suggest that understanding the details
will require extremely accurate ($\sim$10\% in temperature) simulation
results.  Such accuracies will require a focused V\&V effort and I
will outline a basic approach for this problem.  Validation requires
an observational constraint similar to the problem we are solving and
we will discuss the potential of hydrogen deficient stars (R Coronae
Borealis stars) as a validation test for type Ia progenitors.

\section{Status of Current Simulations}

A great deal of work has already been done studying the merger of
white dwarfs (e.g. Mochkovitch \& Livio 1989, 1990; Benz et al. 1990;
Segretain et al. 1997, Guerrero et al. 2004, Yoon et al. 2007).  Let's
focus on the work of the last two papers.  The Guerrero et al. (2004)
work studied a series of binary systems, with range of masses for the
binary components: (0.4,0.4\,M$_\odot$), (0.4,0.6\,M$_\odot$),
(0.4,1.2\,M$_\odot$), (0.6,0.8\,M$_\odot$), (0.6,1.0\,M$_\odot$),
(1.0,1.2\,M$_\odot$).  In all cases, the systems merged after a few
orbital periods, or a few hundred seconds, corresponding to mass
transfer rates of nearly $10^{-2}$\,M$_\odot$\,s$^{-1}$.  The white
dwarf can't incorporate this material on so such a timescale (it is
limited to the Eddington accretion rate: $\sim 10^{-5} {\rm M_\odot
yr^{-1}}$), so most of this material initial builds an atmosphere
around the white dwarf.  This material then accretes at the Eddington
rate.  Such high accretion rates would, using the Nomoto \& Kondo
analysis, ultimately collapse to form AICs.

But Saio \& Nomoto (2004) and Yoon et al. (2007) have found that 
not all such systems need necessarily produce AICs.  If the white 
dwarf is differentially rotating, we can expect very different results.  
Yoon et al. (2007) found that if the white dwarf is spinning fast enough, 
core contraction can be delayed until the core heats up sufficiently to 
ignite at low densities, driving a thermonuclear explosion and a type Ia 
supernova.  Unfortunately, their results were not sufficiently accurate 
to determine exactly which fate each system would follow.

One last set of results has thrown yet another wrench into our
current understanding of double degenerate mergers.  Motl et
al. (2002) and D'Souza et al. (2006) found a very different result for
the fate of a binary system with mass ratios similar to those used by
Guerrero et al. (2004).  Their work found that, instead of the
catastrophic destruction of the mass-losing star, the mass transfer is
stable.  If the mass transfer is stable, the accretion rate would then
be determined by the coalescence time from gravitational wave
radiation:
\begin{equation}
T_{\rm merger} = 5\times10^5 (A/10^{10}{\rm cm})^4(M_\odot/M_1)(M_\odot/M_2)(M_\odot/M_{\rm tot}){\rm yr},
\end{equation}
where $A$ is the orbital separation, $M_1,M_2,M_{\rm tot}$ are,
respectively, the masses of the primary, secondary and total binary
mass of the system.  For typical separations of a few times
10$^9$\,cm, the mass transfer timescale is on par with the Eddington
rate.  If we need to solve the temperature evolution of the accreting
matter to better than 10\%, such differences (10 orders of magnitude -
although the corresponding error in the temperature is probably less
than a factor of 2) in the mass transfer rate will make a large
difference.

\section{What do we expect?}

If we require 10\% accuracies in the temperature, we need to take
numerical testing to a new level.  We must understand the deficiencies
and strengths of each code and we must track down the differences in
the simulations.  D'Souza et al. (2006) use a very different code than
that used in all other studies.  Their code is grid, not particle,
based.  The disadvantage of such a technique is that grid-based codes
have trouble conserving angular momentum, but this team has worked
very hard to remove this issue from their simulations.  An advantage
of this scheme is that it can easily model low mass-transfer rates,
but our preliminary SPH calculations have shown that with the SPH
particle counts we can afford today (1-10 million particles), modeling
low mass transfer rates becomes more tractable.  Also, grid codes model
shocks differently than SPH codes (performing better on shock-tube
problems where the shock is along the grid).  Currently, the
D'Souza et al. (2006) result does not include shocks.

\begin{figure}[t]
\plottwo{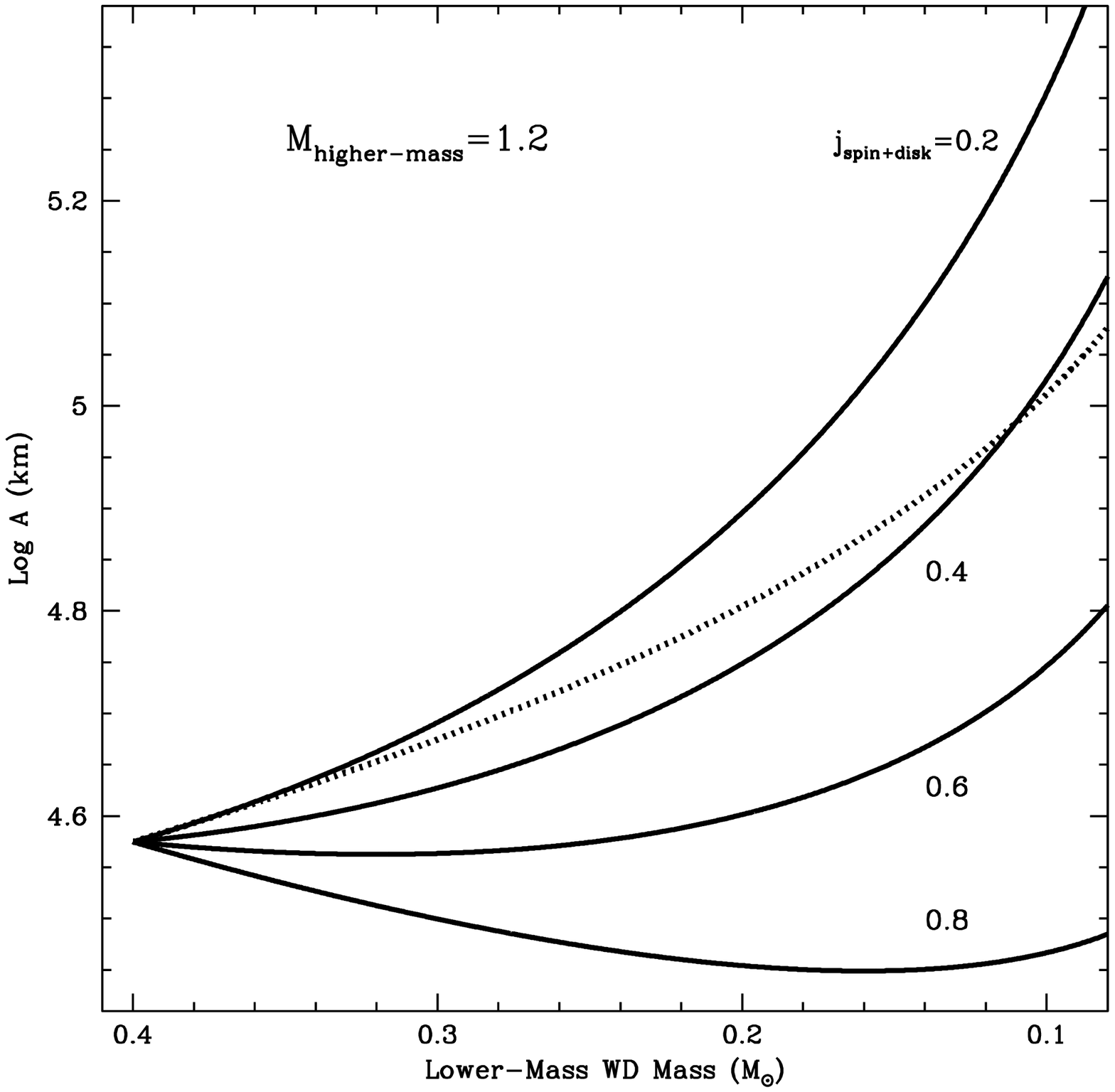}{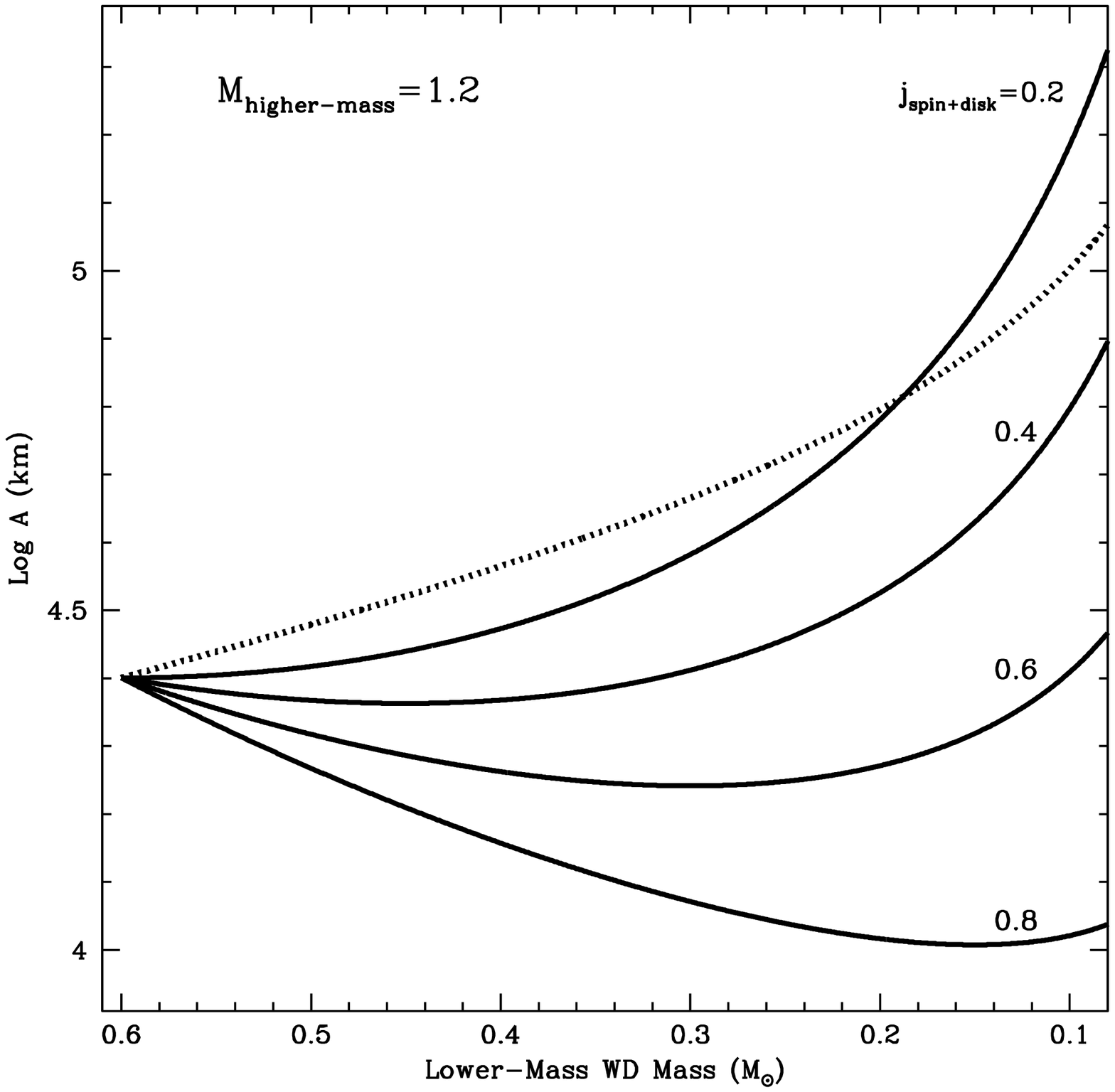}
\caption{Semi-analytic estimate of the evolution of the orbital
separation (solid line) and separation where the lower-mass white
dwarf overfills its Roche lobe (dashed line).  As the white dwarf
loses mass (left to right), it expands and the orbital separation
where it will overfill its Roche lobe also increases.  Depending 
upon how much orbital angular momentum is lost in the mass transfer 
phase (and depending upon the relative masses of the two stars), 
the actual orbital separation may decrease or increase.  To 
understand this graph, let's study one or two possible tracks.    
If $j_{\rm spin+disk}=0.2$ for the 0.4,1.2\,M$_\odot$ star 
merger (left panel), the orbital separation will increase faster 
than the white dwarf expands:  fate - steady mass transfer.  
If $j_{\rm spin+disk}=0.4$ for the 0.4,1.2\,M$_\odot$ star 
merger (left panel), the white dwarf will expand faster 
than the orbit until the mass of the white dwarf falls below 
0.14\,M$_\odot$:  fate - runaway accretion.}
\end{figure}

\begin{figure}
\plotfiddle{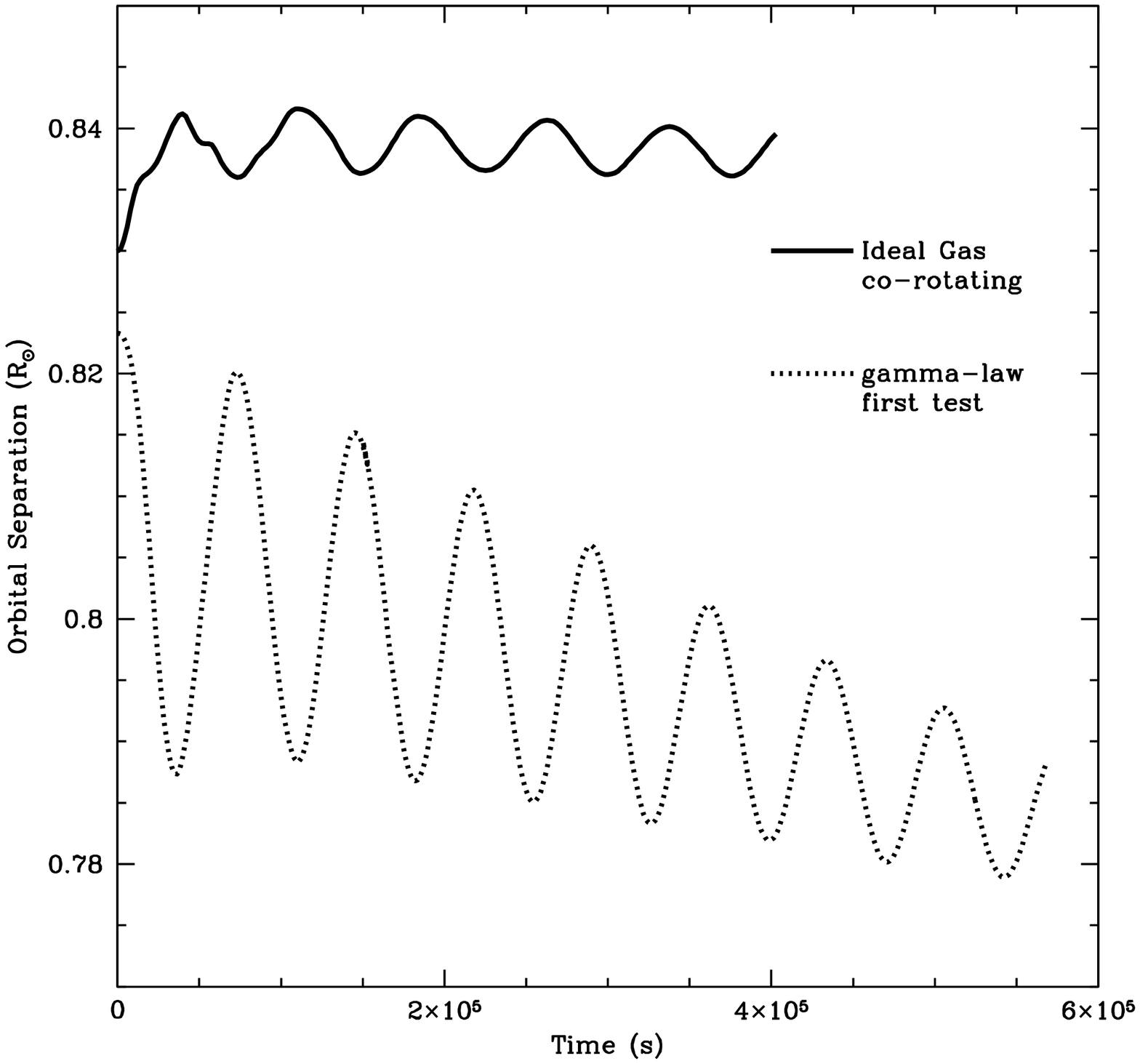}{2.8in}{0.}{45}{45}{-155}{-90}
\caption{Orbital separation as a function of time for our two runs:
co-rotating with shocks (solid), no co-rotation without shocks
(dashed).  The oscillations occur because the system is not in 
perfect circular orbits.  The decrease in the orbital separation 
in the case of our non-corotating case is primarily due to the 
fact that orbital angular momentum is converted into spin angular 
momentum in the stars.}
\end{figure}

What do we expect the result to be from analytic (or semi-analytic)
estimates?  When gravitational radiation brings together the two white
dwarfs in a double degenerate merger, the lower-mass white dwarf
overfills its Roche lobe and accretes onto the higher-mass white
dwarf.  The subsequent accretion is affected by two processes.  First,
because the lower-mass white dwarf is supported by degeneracy
pressure, it will expand as it loses mass.  If the orbital separation
were kept constant, the accretion process would quickly run away and
the white dwarf would be completely disrupted in a few orbits.  The
Second effect is the fact that, if orbital angular momentum is
conserved, the orbital separation will increase as the lower-mass star
accretes onto its higher-mass companion.  This effect would try to
push the system into stable accretion.  In reality, orbital angular
momentum is not strictly conserved, but it is still likely that the
orbit will expand during the accretion process.  It is then the
competition between the expansion of the white dwarf pushing toward
runaway accretion and the expansion of the orbit pushing toward 
stable accretion that drives the evolution of the accretion.

Clayton et al. (2007) provide a more quantitative analysis of this 
processs.  To estimate the expansion of the white dwarf, they used 
the following formula for the white dwarf radius (Nauenberg 1972):  
\begin{equation}
R_{\rm WD} = 10^4 (M_{\rm WD}/0.7M_\odot)^{-1/3} (1-M_{\rm WD}/M_{\rm
CH})^{1/2}(\mu_e/2)^{-5/3} {\rm km},
\end{equation}
where $M_{\rm WD}$ is the white dwarf mass, $M_{\rm CH}$ is the
Chandrasekhar mass and $\mu_e$ is the mean molecular weight per
electron of the white dwarf.  For the evolution of the orbital
separation, Clayton et al. (2007) used (Podsiadlowski et al. 1992;
Fryer et al. 1999):
\begin{equation}
A/A_0 = (M_{\rm low-mass}/M_{\rm low-mass}^0)^{C_1}(M_{\rm
high-mass}/M_{\rm high-mass}^0)^{C_2}
\end{equation}
where $A_0$, $M_{\rm low-mass}^0$, $M_{\rm high-mass}^0$ are the
initial values for the orbital separation and masses of the lower and
higher mass white dwarfs.  The angular momentum conservation or lack
thereof is including in two coeeficients: $C_1 \equiv -2 + 2 j_{\rm
disk+spin}$ and $C_2 \equiv -2 - 2 j_{\rm disk+spin}$ where $j_{\rm
disk+spin}$ is the term for the specific angular momentum of the
accreted material that is lost to either spinning up the companion or
to an accretion disk (see Fryer et al. 1999 for details).

Figure 1 shows the competition between these two effects for two
different binary systems: 0.4,1.2\,M$_\odot$ components and
0.6,0.9\,M$_\odot$ components.  The orbtial separation where the
lower-mass white dwarf overfills its Roche-lobe expands as it loses
mass.  The evolution of the orbital separation depends upon our value
of $j_{\rm disk+spin}$.  Of course, the actual fate of a merging system 
will depend upon a more exact representation of the white dwarf radius, 
but the primary uncertainty in the fate depends upon the determination 
of the value of $j_{\rm disk+spin}$.  The differing results between 
D'Souza et al. (2006) and the SPH calculations all probably reside 
in different values for this quantity.  

\section{On the Path to Predictive Simulations}

Code comparison has long been used effectively in astrophysics to
estimate theoretical errors in a numerical solution.  If we can
understand the differences in the simulations, we can not only
estimate the numerical errors, but we can also find ways to improve
the codes and minimize these errors.  Here is a first attempt 
at understanding these differences.   

To help better explain these differences, we produce some of our own
calculations using the SNSPH code (Fryer et al. 2006).  The SPH
algorithm in this code copies that used by Fryer et al. (1999) and has
already been used on a number of binary calculations (e.g. Fryer \&
Heger 2005).  We present 2 preliminary calculations both using the
initial setup from D'Souza et al. (2006) with a mass ratio of 0.4.
The first simulation uses the same polytropic equation of state from
D'Souza et al. (2006).  With this equation of state, shocks do not
occur.  In this calculation, we did not initially put the binary
components in co-rotation.  In the second simulation we use an ideal
gas equation of state to include the effects of shocks.  In this
simulation, we placed the stars in co-rotation prior to starting the
simulation.

Our initial orbital separation places the binaries close enough that
the lower-mass star overfills its Roche radius and is accreting onto
the more massive star.  Figure 2 shows the time evolution of this
orbital separation.  In the first simulation, the fact that the stars
were not co-rotating meant that after about 10 orbits, the orbital
separation decreased by nearly 2\%.  This star will ultimately be
disrupted in less than a 20-30 orbits.  This timescale is longer than
most past calculations of WD mergers, but much shorter than that
predicted by D'Souza et al. (2006).

\begin{figure}
\plotone{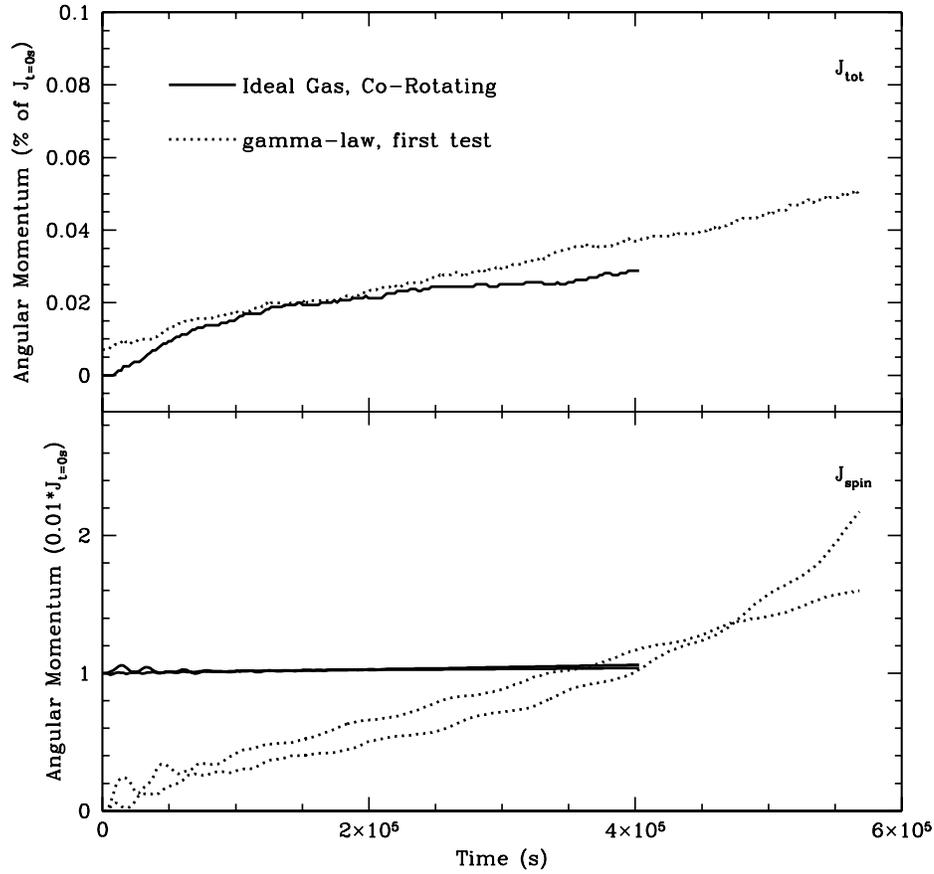}
\caption{top:  Total angular momentum in the binary system as a 
function of time.  bottom:  angular momentum in the spin of the 
stars (as a fraction of total angular momentum) as a function of 
time.  In both our simulations, numerical artifacts which we have 
yet to determine allow the angular momentum to increase 0.02\% 
over about 10 orbits.  But the biggest effect on the angular 
momentum is the conversion of orbital angular momentum to spin 
angular momentum in the case of our stars that are initially 
non co-rotating.  For the co-rotating stars, the stars do 
not extract much angular momentum.}
\end{figure}

Why are we getting a result that lies in between these two extremes.
If we look more carefully at the angular momentum (Fig. 3), we see
that it increases as a function of time.  The total angular momentum
in our simulations has increased by 0.02\% after 10 orbits.  This
numerical artifact should cause the orbital separation to increase.
So why does it decrease in the case of the non co-rotating run?  The
answer lies in the amount of orbital angular momentum converted into
the spin of the angular momentum.  The lower panel shows the total
amount of the angular momentum in stellar spin angular momentum. In
the non co-rotating case, the spin angular momentum increases with
time.  This angular momentum is being taken from the orbital angular
momentum.  Although the total angular momentum is increasing at the
0.02\% level over 10 orbits, over 3\% of the orbital angular momentum
is being converted into spin angular momentum at the same time,
causing a net loss in the orbital angular momentum and forcing the
binary to coalesce.  Note that in our co-rotating initial conditions,
very little orbital angular momentum is converted into spin angular
momentum.

\begin{figure}
\plotone{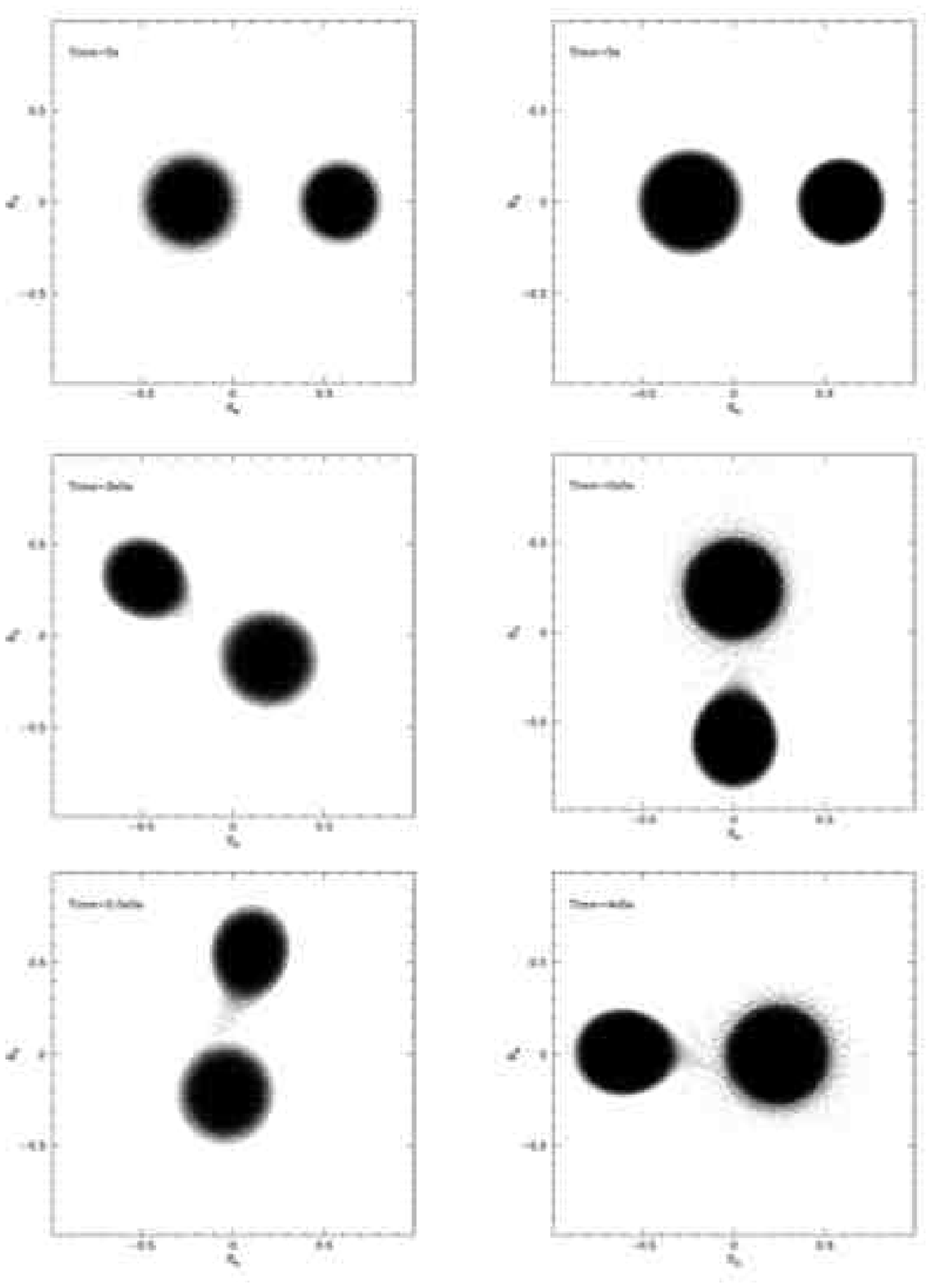}
\caption{Three snapshots in time for both our ``no-shocks'' simulation 
(left column) and ``shocked'' simulation (right column).  In the 
no-shock case, all of the matter accretes directly onto the 
accretor.  In the shocked case, an atmosphere builds up around 
the accreting star that ultimately envelopes both stars.}
\end{figure}

So it seems that when we are careful about our initial conditions, the
merger time is very much longer than the timescale predicted by all of
the previous SPH calculations.  This is probably because most of the
past work did not worry too much about the initial conditions.  But
does this mean that as we remove numerical artifacts we will
ultimately reach the result of D'Souza et al. (2006)?  Not quite.
Recall that D'Souza et al. (2006) did not include shock heating.
Figure 4 shows the evolution of matter for both our simulations.  In
the unshocked simulation, the material is immediately incorporated
into the accreting star.  Since the star is co-rotating, the material
incorporated into the white dwarf must have given its angular momentum
back to the orbit and it is unlikely that much angular momentum is
lost at all form the binary system.  But in the shocked simulation,
the accreting material forms an atmosphere around the entire binary
system.  First, this material must take angular momentum from the orbit.

We now can understand the stable accretion in the D'Souza et al. (2006) 
results and why that may not be the right solution either.  Without shocks, 
the D'Souza et al. (2006) essentially guaranteed that the value for
$j_{\rm disk+spin}$ was set to 0 because it allows the accreting
object to incorporate all of the accreting material.  With shocks, an
atmosphere forms and $j_{\rm disk+spin}$ is definitely more than 0.
The D'Souza et al. (2006) result is also not the final answer.  We
simply have yet to converge on the exact value for this factor, and
without it, we can not know the mass transfer rate.  And we definitely
can not predict accurate temperature profiles for the accreting matter
or the accreting white dwarf.

\section{Summary}

So what can we say at this point in time?  First, no calculation has
yet accurately calculated the mass transfer rate in double degenerate
mergers.  We know that it is probably not as fast as that shown in
most SPH calculations of type Ia progenitors, but it is probably
faster than those results using unshocked matter.  If we only had to
distinguish between a mass transfer rate above or below the Nomoto \&
Kondo (1991) line, these simulations would be sufficient: the mass
transfer rate is almost certainly above this line, possible even above
the Eddington accretion rate (that means it will form a thick disk).
But if we want to get accurate temperature evolution profiles, we will
have to do more accurate calculations.

\begin{figure}
\plotone{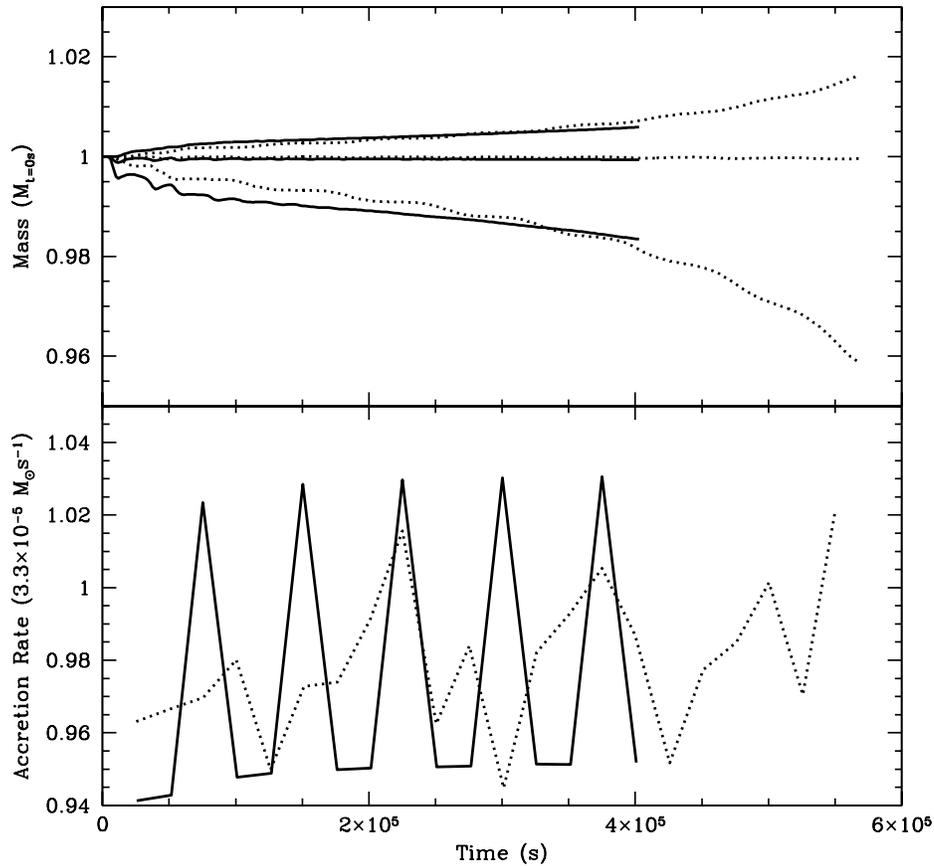}
\caption{Star masses and accretion rates as a function of time.}
\end{figure}

Figure 5 shows the accretion rates for our current simulations.
Although we have much more testing to do before such quantities like
accretion rate can be trusted, we can still study a number of trends
that may have implications in the observations to help constrain the
simulations.  For example, the accretion is periodic caused by slight
errors in the initial orbit.  How does this alter the nuclear burning?  It
will affect the temperature as well as the mixing.  We may be able to
use observations of R Coronae Borealis stars to place limits on the
level of eccentricity in nature.

Such high-precision calculations can not be done without systematic
verification and validation.  Here we have shown some of the tests
that can be used in V\&V: comparing (and understanding) the differences
in code results in concert with analytic comparisons.  This requires a
step-by-step process, adding physics one piece at a time to understand
its effect.  Convergence studies will also prove useful in this
problem.  This all fits under the ``verification'' of a code.  It would 
be nice to also conduct some validation tests - comparison to a very 
similar problem where the data is more plentiful.  It could well be 
that stars like R Coronae Borealis stars provide such a validation 
experiment.

\acknowledgements This work is almost entirely driven by encouragement 
and helpful conversations with Geoff Clayton, Joel Tohline, Falk Herwig 
and Orsola DeMarco.  It is funded by an exploratory research grant at 
Los Alamos National Laboratory.


\end{document}